\documentclass{article}
\usepackage[centertags]{amsmath}
\usepackage{amsfonts}
\usepackage{amssymb}
\usepackage{amsthm}
\usepackage{newlfont}
\usepackage{graphicx}
\newcommand \qq {\qquad}
\newcommand \q  {\quad}
\theoremstyle{plain}

\theoremstyle{remark}

\theoremstyle{example}

\numberwithin{equation}{section}

\begin{document}
 {\large
 \centerline{\bf\Large Potential symmetry and invariant solutions of
 Fokker-Planck equation in}
 \vskip0.3cm
 \centerline{\bf\Large cylindrical coordinates related to magnetic
 field diffusion }
\vskip0.3cm
  \centerline{\bf\Large in magnetohydrodynamics including the Hall current}
 \vskip0.5cm
\centerline{\Large A. H. Khater$^{a,1}$, D. K. Callebaut$^b$, S. F.
Abdul-Aziz$^a$, } \vskip0.3cm
 \centerline{\Large T. N. Abdelhameed$^a$}
  \vskip0.5cm
 \centerline{\large  $^a$Departement of Mathematics, Faculty of Science,
  Beni-Suef University, Beni-Suef, Egypt}
 \centerline{\large $^b$Departement Natuurkunde, CDE, University of Antwerp, B-2610 Antwerp, Belgium}
 \vskip1.5cm
 \noindent
{\bf\large Abstract}
 \noindent

Lie groups involving potential symmetries are applied in connection
with the system of magnetohydrodynamic
equations for incompressible matter with Ohm's law for finite
resistivity and Hall current in cylindrical geometry.
Some simplifications allow to obtain a Fokker-Planck type
equation. Invariant solutions are
obtained involving the effects of time-dependent flow and the
Hall-current. Some interesting side results of this approach are
new exact solutions that do not seem to have been reported in the
literature.

\footnotetext[1]{Corresponding author A. H. Khater, e-mail
khater$_{-}$ah@yahoo.com}

\newpage
 \section{Introduction}

 \noindent

  Recently, Khater  et al. [1, 2]  have analyzed the
generalized one-dimensional Fokker-Planck (FP) equation and the
inhomogeneous NL diffusion equation through the application of the
potential symmetries. For a brief exposition of the potential
symmetries and of the equations of magnetohydrodynamics (MHD): see
[2].

 \noindent
 Some interesting side results of the present study are new
exact solutions that do not seem to have been reported in the
literature.

 \noindent
 This paper is organized as follows:

 \noindent
Section 2 deals with the determination of the potential
symmetries. In section 3, we analyze the invariant solutions of
the MHD equations for various cases corresponding to physically
interesting situations. Section 4 gives the conclusions.

  \section{Determination of the potential symmetries}

 \noindent

Consider a partial differential equation (PDE), $R$, of order m
written in a conserved form: ( [3] and references therein)
\begin{equation}
\sum^{n}_{i=1}\frac{\partial}{\partial
x_{i}}F_{i}[x,u,u_{1},u_{2},...,u_{m-1}]=0
\end{equation}
with  $n\geq 2$ independent variables $x=(x_{1},x_{2},...,x_{n})$
and a single dependent variable $u$. For simplicity, we consider a
single PDE - the generalization to a system of PDEs in a conserved
form is straight-forward. The indexes of $u$ indicate the order of
the derivative. If a given PDE is not written in a conserved form,
there are a number of ways of attempting to put it in a conserved
form. These include a change of variables (dependent as well as
independent), an application of Noether's theorem [4], direct
construction of conservation laws from field equations [5], and some
combinations of them.

Using some simplifications [2] we may put the equation of the
evolution of flow in the MHD system for cylindrical coordinates,
which is a generalized FP equation, in the following conservative
form:
\begin{equation}
u_{t}-(\frac{1}{r^{2}}u_{\theta}+\frac{1}{r}\lambda_{t}
\theta u)_{\theta}=0
\end{equation}
with $\lambda$ a function of $r$ and $t$; By considering a potential $v$ as an auxiliary
unknown function, the following system $S$ can be associated with
(2.2):
\begin{equation}
v_{\theta}=u \q ,\q
v_{t}=\frac{1}{r^{2}}u_{\theta}+\frac{1}{r}\lambda_{t} \theta u.
\end{equation}

\noindent
 It is well known that the homogeneous linear system,
  which characterizes the generators, is  obtained  from [6]
\begin{equation}
Y^{(1)}(v_{\theta}-u)|_{s}=0 \q , \q
Y^{(1)}(v_{t}-\frac{1}{r^{2}}u_{\theta}-\frac{1}{r}\lambda_{t}
 \theta u)|_{s}=0
\end{equation}
which must hold identically.

\noindent
 Here, $Y^{(1)}$ is the operator:
\begin{equation}
Y^{(1)}=\tau \frac{\partial}{\partial t}+\xi
\frac{\partial}{\partial \theta}+\eta \frac{\partial}{\partial
u}+\phi \frac{\partial}{\partial
v}+\eta^{(1)}_{1}\frac{\partial}{\partial
u_{t}}+\eta^{(1)}_{2}\frac{\partial}{\partial
u_{\theta}}+\phi^{(1)}_{1}\frac{\partial}{\partial
v_{t}}+\phi^{(1)}_{2}\frac{\partial}{\partial v_{\theta}};
\end{equation}
\[
\begin{split}
\eta^{(1)}_{1}&=\eta_{t}+(\eta_{u}-\tau_{t})u_{t}-\tau_{u}u^{2}_{t}-
\xi_{t}u_{\theta}-\xi_{u}u_{t}u_{\theta}+\eta_{v}v_{t}-\tau_{v}u_{t}v_{t}
-\xi_{v}u_{\theta}v_{t},\\
 \eta^{(1)}_{2}&=\eta_{\theta}+(\eta_{u}-\xi_{\theta})u_{\theta}-
\tau_{\theta}u_{t}-
\tau_{u}u_{\theta}u_{t}-\xi_{u}u^{2}_{\theta}+\eta_{v}v_{\theta}-
\tau_{v}u_{t}v_{\theta}-\xi_{v}u_{\theta}v_{\theta},\\
 \phi^{(1)}_{1}&=\phi_{t}+(\phi_{v}-\tau_{t})v_{t}-\tau_{v}v^{2}_{t}+
\phi_{u}u_{t}-\tau_{u}u_{t}v_{t}-\xi_{t}v_{\theta}-\xi_{u}u_{t}v_{\theta}
-\xi_{v}v_{t}v_{\theta},\\
 \phi^{(1)}_{2}&=\phi_{\theta}+(\phi_{v}-\xi_{\theta})v_{\theta}
-\xi_{v}v^{2}_{\theta}+\phi_{u}u_{\theta}-\tau_{\theta}v_{t}
-\tau_{u}u_{\theta}v_{t}-\tau_{v}v_{\theta}v_{t}-\xi_{u}u_{\theta}v_{\theta}
\end{split}
\]
Eq. (2.4) becomes
\begin{equation}
\phi_{\theta}+(\phi_{v}-\xi_{\theta})v_{\theta}-\xi_{v}v^{2}_{\theta}+
\phi_{u}u_{\theta}-\tau_{\theta}v_{t}-\tau_{u}u_{\theta}v_{t}-\tau_{v}v_{\theta}
v_{t}-\xi_{u}u_{\theta}v_{\theta}-\eta =0,
\end{equation}
\begin{equation}
\begin{split}
\phi_{t}&+(\phi_{v}-\tau_{t})v_{t}-\tau_{v}v^{2}_{t}+\phi_{u}u_{t}
-\tau_{u}u_{t}v_{t}-\xi_{t}v_{\theta}-\xi_{u}u_{t}v_{\theta}\\
&-\xi_{v}v_{t}v_{\theta} -\frac{1}{r}\lambda_{t}\theta \eta
-\frac{1}{r}\lambda_{t}u\xi-\frac{1}{r^{2}}[\eta_{\theta}+(\eta_{u}-\xi_{\theta})
u_{\theta}-\tau_{\theta}u_{t}\\
&-\tau_{u}u_{\theta}u_{t}-\xi_{u}u^{2}_{\theta}+
\eta_{v}v_{\theta}-\tau_{v}u_{t}v_{\theta}-\xi_{v}u_{\theta}
v_{\theta}]-\frac{1}{r}\theta \lambda_{tt}\tau u=0
\end{split}
\end{equation}
On substituting $v_{\theta}$ by $u$, and $v_{t}$ by
$\frac{1}{r^{2}}u_{\theta}+\frac{1}{r}\lambda_{t}\theta u$ in
Eqs. (2.6) and (2.7), we get:
\begin{equation}
\tau =\tau (t) \q , \q \xi=\xi (\theta ,t) \q ,\q \phi_{u}=0
\end{equation}
\begin{equation}
\phi_{\theta}-\eta +(\phi_{v}-\xi_{\theta})u=0
\end{equation}
\begin{equation}
\phi_{v}-\eta_{u}-\tau_{t}+\xi_{\theta}=0
\end{equation}
\begin{equation}
\phi_{t}-\frac{1}{r^{2}}\eta_{\theta}-\frac{1}{r}\lambda_{t}\theta\eta+
[\frac{1}{r}\lambda_{t}\theta
(\phi_{v}-\tau_{t})-\frac{1}{r}\lambda_{t}\xi
-\xi_{t}-\frac{1}{r^{2}}\eta_{v}+\frac{1}{r}\lambda_{tt}\theta
\tau]u=0;
\end{equation}
with
\begin{equation}
\eta=f(\theta ,t)u+g(\theta ,t)v \q ,\q \phi =k(\theta ,t)v,
\end{equation}
where $f,g$ and $k$ are arbitrary smooth functions of $\theta$ and $t$.

\noindent
  On solving the above system of  Eqs. (2.8)-(2.12), we get:
\begin{equation}
\tau =\tau (t) \q ,\q \xi =\xi (\theta ,t) ,
\end{equation}
\begin{equation}
k_{\theta}-g =0,
\end{equation}
\begin{equation}
K-f-\xi_{\theta}=0,
\end{equation}
\begin{equation}
2\xi_{\theta}-\tau_{t}=0
\end{equation}
\begin{equation}
k_{t}-\frac{1}{r^{2}}g_{\theta}-\frac{1}{r}\lambda_{t}\theta g
=0 ,
\end{equation}
\begin{equation}
\frac{1}{r}\lambda_{t}\theta \tau
+\frac{1}{r^{2}}g+\frac{1}{r^{2}}f_{\theta}+\xi_{t}+\frac{1}{r}\theta
\lambda_{t}\xi_{\theta}+\frac{1}{r}\lambda_{t}\xi=0 ,
\end{equation}
\begin{equation}
g_{t}=(\frac{1}{r^{2}}g_{\theta}+\frac{1}{r}\lambda{t}\theta
g)_{\theta}.
\end{equation}
In solving the above system of Eqs.(2.13)-(2.19), we confine our
attention to physically interesting situations.

\section{Invariant solutions}

 \noindent

From now on, we will denote by $c_{0}-c_{13}$ arbitrary
constants

\noindent
 Let $\lambda =\frac{1}{2r}$

\noindent
 In this  case, the  infinitesimal  symmetries  are  given  by :
\begin{equation}
\left.
\begin{split}
\tau&=-2r^{2}c_{4}e^{-\frac{t}{r^{2}}}+c_{5},\\
\xi&=c_{4}e^{-\frac{t}{r^{2}}}\theta +c_{3}e^{-\frac{t}{2r^{2}}}-
  2\sqrt{2r}c_{0}e^{\frac{t}{2r^{2}}},\\
\eta&=(\sqrt{2r}c_{0}e^{\frac{t}{2r^{2}}} \theta +c_{1}-c_{2})u+
 \sqrt{2r}c_{0}e^{\frac{t}{2r^{2}}}v,\\
\phi&=(\sqrt{2r}c_{0}e^{\frac{t}{2r^{2}}}\theta +c_{1})v
\end{split}
\right\}
\end{equation}
\noindent
 Then, we obtain point symmetries with the
following  generators :
\[
\begin{split}
Y_{1}&:\tau=0, \q \xi=-2\sqrt{2r}e^{\frac{t}{2r^{2}}}, \q
 \eta =\sqrt{2r}e^{\frac{t}{2r^{2}}}\theta u +\sqrt{2r}e^{\frac{t}{2r^{2}}}v, \q
  \phi =\sqrt{2r}e^{\frac{t}{2r^{2}}}\theta v ,\\
Y_{2}&:\tau =\xi =0, \q \eta =u , \q \phi =v ,\\
Y_{3}& \tau =\xi =\phi =0 , \q \eta =-u ,\\
 Y_{4}&: \tau = \eta =\phi = 0, \q \xi =e^{-\frac{t}{r^{2}}} \\
 Y_{5}&: \tau =-2r^{2}e^{-\frac{t}{r^{2}}}, \q \xi =e^{-\frac{t}{r^{2}}}\theta ,
 \q \eta =\phi =0, \\
 Y_{6}&:\tau =1, \q \eta =\phi =\xi =0
\end{split}
\]
and $\infty$-dimensional symmetry, which is a consequence of the
linearity [7]. It is clear that, $Y_{1}$ is only a potential
symmetry for Eq.(2.2).

\noindent
 For the potential symmetry $Y_{1}$, the characteristic
system related to the invariant surface conditions reads:
\begin{equation}
v=c_{7}e^{\frac{-\theta^{2}}{4}}, \qq t=c_{6}
\end{equation}
\begin{equation}
u=(c_{8}-\frac{c_{7}}{2}\theta)e^{\frac{-\theta^{2}}{4}}
\end{equation}
If we assume $t=c_{6}=z$ as a parameter, $c_{7}=h_{2}(z)$, and
$c_{8}=h_{1}(z)$ in Eqs. (3.2) and (3.3), we obtain:
\begin{subequations}
\begin{gather}
u=(h_{1}(z)-\frac{h_{2}(z)}{2}\theta)e^{\frac{-\theta^{2}}{4}},
\end{gather}
\begin{gather}
v=h_{2}(z)e^{\frac{-\theta^{2}}{4}}; \q z=t
\end{gather}
\end{subequations}
Now, to find the solutions $F^{*}_{E}$, we introduce Eq. (3.4a) in
Eq. (2.2) obtaining:
\begin{equation}
h^{'}_{1}-\theta (\frac{h^{'}_{2}}{2}+\frac{h_{2}}{4r^{2}})=0
\end{equation}
which must hold for any value of $\theta$.

\noindent
 From Eq. (3.5), we have the system $\overset=\varphi$ as:
\begin{equation}
\left.
\begin{split}
h^{'}_{1}&=0,\\
\frac{h^{'}_{2}}{2}+\frac{h_{2}}{4r^{2}}&=0
\end{split}
\right\}
\end{equation}
which on solving, yields
\begin{equation}
\left.
\begin{split}
h_{1}(z)&=c_{9} ,\\
h_{2}(z)&=c_{10}e^{-\frac{t}{2r^{2}}}
\end{split}
\right\}
\end{equation}
Then, the family $F^{*}_{E}$  is therefore:
\begin{equation}
u=(c_{9}-\frac{c_{10}}{2}\theta
e^{-\frac{t}{2r^{2}}})e^{-\frac{\theta^{2}}{4}}
\end{equation}
Also, Eq. (3.4a) is a family of solutions of the first-order
equation:

\noindent
 To find the solutions $F_{E}$, we introduce Eq. (3.4) in Eq.
(2.3) obtaining the system $\overset=\varphi$ as:
\begin{equation}
\left.
\begin{split}
h_{1}&=0,\\
h^{'}_{2}+\frac{h_{2}}{2r^{2}}&=0
\end{split}
\right\}
\end{equation}
which on solving, yields
\begin{equation}
\left.
\begin{split}
h_{1}(z)&=0,\\
h_{2}(z)&=c_{11}e^{-\frac{t}{2r^{2}}}
\end{split}
\right\}
\end{equation}
Then, the family $F_{E}$ is therefore:
\begin{equation}
u=-\frac{c_{11}}{2}\theta
e^{-\frac{1}{4r^{2}}(\theta^{2}r^{2}+2t)}\q(\text{see fig. (2)}).
\end{equation}
It is clear that, $F_{E}$ is enclosed in $F^{*}_{E}$, which are
new solutions as far as we know.

 \vskip0.5cm
\noindent
 {\bf\large Particular case. }

\noindent If, $f=\lambda_{t}\theta$, $g=v^{z}=0$,
and $u=\Theta_{\theta}$ in Eqs. (2.10)-(2.13) we obtain that
\begin{equation}
f_{1}(t)=0,
\end{equation}
\begin{equation}
u=u(t)
\end{equation}
\begin{equation}
u_{t}=\frac{1}{r}\lambda_{t}u,
\end{equation}
\begin{equation}
\lambda_{tt}+\frac{2}{r}\lambda^{2}-\frac{1}{r^{2}}\lambda =0
\qq \upsilon_{m}=1
\end{equation}
Solving Eq. (3.15), yields
\begin{equation}
\lambda_{t}=\frac{1}{2r+c_{12}e^{-\frac{t}{r^{2}}}}
\end{equation}
Then, the family $F^{*}_{E}$ is given by:
\begin{equation}
u=c_{13}\sqrt{2re^{\frac{t}{r^{2}}}+c_{12}}\q(\text{see fig.
(3)}).
\end{equation}
It is clear that, $F_{E}$ is enclosed in $F^{*}_{E}$, which are
new solutions as far as we know.
  \section{Conclusion}

 \noindent

    In this paper, we made an analysis for the FP-type equation
    with convection given by the plasma flow with finite electrical
 conductivity and Hall current. This method based on potential symmetries turns out to be an alternative, systematic and powerful technique for the determination of the solutions of linear or nonlinear PDEs,
     single or a system. The infinitesimals, similarity variables,
     dependent variables, and reduction to quadrature or exact solutions of
     the mentioned FP-type equation (in cylindrical coordinates) for physically
     realizable forms of $\lambda $, and $q$ are also obtained.

\noindent
    The similarity solutions given here do not seem to have been reported in
    the literature. Some of these solutions are unbounded. However, one can deal
     with them as various methods have
    been elaborated to analyze the properties of unbounded (particularly explosive type)
    solutions of the Cauchy problem of quasilinear parabolic equations of type (2.2).

\newpage
\noindent
 {\bf\large References}
\begin{description}
\item{[1]} A. H. Khater, M. H. M. Moussa, S. F. Abdul-Aziz, Physica
               A 304  (2002) 395.
    \item{[2]} A. H. Khater, D. K. Callebaut, S. F. Abdul-Aziz, T. N. Abdelhameed, Physica
               A., 341 (2004) 107.
  \item{[3]} G. W. Bluman, S. Kumei, G.J. Reid, J. Math. Phys. 29
             (1988) 806.
\item{[4]} J. D. Logan, Invariant Variational Principles, Academic Press,
               New  York, 1977.
               \item{[5]} S. C. Anco, G. W. Bluman, Phys. Rev. Lett. 78 (1997) 2869.

    \item{[6]} G. W. Bluman, J. D. Cole, Similarity Methods for Differential
          Equations (Springer-Verlag, New York, 1974).

    \item{[7]} G. W. Bluman, S. Kumei, Symmetries and Differential
             Equations, Springer-Verlag, New York, 1989.

    \end{description}

\newpage
\noindent {\bf Figure Captions :}\\

\noindent {\bf Fig. (1a):} The magnetic field  in the surface with
$\mu_{1}=10, \mu_{2}=0.1$ and $m=0.001$\\

\noindent {\bf Fig. (1b):} The magnetic field in the
surface with $\mu_{1}=-10, \mu_{2}=0.1$ and $m=0.001$\\

\noindent {\bf Fig. (1c):} The magnetic field in the
surface with $\mu_{1}=10, \mu_{2}=0$ and $m=0.01$\\

\noindent {\bf Fig. (2a):} The solution for a Fokker-Planck in the
surface with $c_{11}=-30$, $t=0$\\

\noindent{\bf Fig. (2b):} The solution for a Fokker-Planck in the
surface with $c_{11}=-30$, $t=1$\\

\noindent{\bf Fig. (2c):} The solution for a Fokker-Planck in the
surface with $c_{11}=30$, $t=1$\\

\noindent{\bf Fig. (3a) :} The solution for a Fokker-Planck in the
surface with (particular case) $c_{13}=5$, $t=0$, $c_{12}=1$\\

\noindent{\bf Fig. (3b) :} The solution for a Fokker-Planck in the
surface with (particular case) $c_{13}=5$, $t=1$, $c_{12}=1$\\

\noindent{\bf Fig. (3c) :} The solution for a Fokker-Planck in the
surface with (particular case) $c_{13}=5$, $t=2$, $c_{12}=-10$\\

 }

\begin{figure}[h]
\begin{center}
\includegraphics[width=60mm,height=60mm]{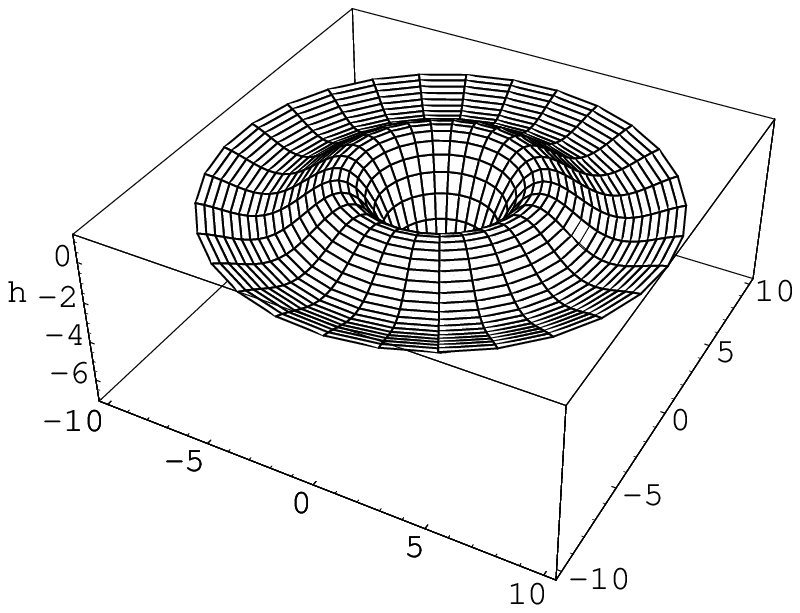}
\end{center}
\caption{fig (1a)}
\end{figure}
\begin{figure}[h]
\begin{center}
\includegraphics[width=60mm,height=60mm]{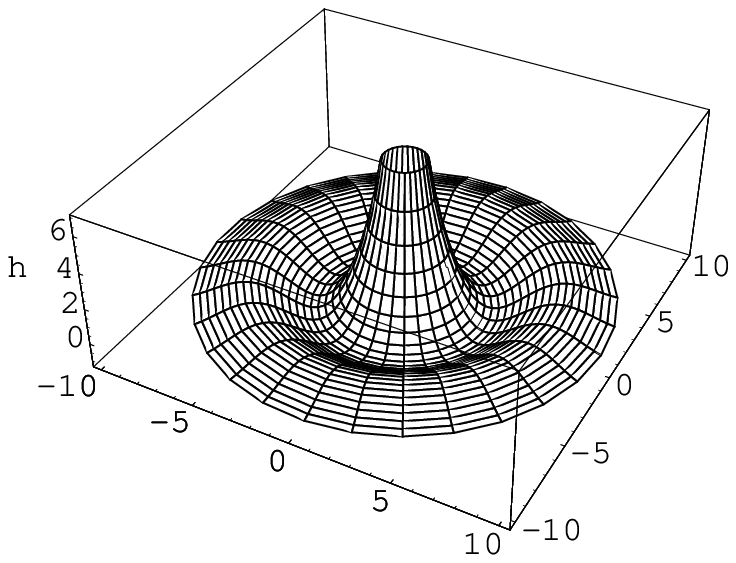}
\end{center}
\caption{fig (1b)}
\end{figure}
\begin{figure}[h]
\begin{center}
\includegraphics[width=60mm,height=60mm]{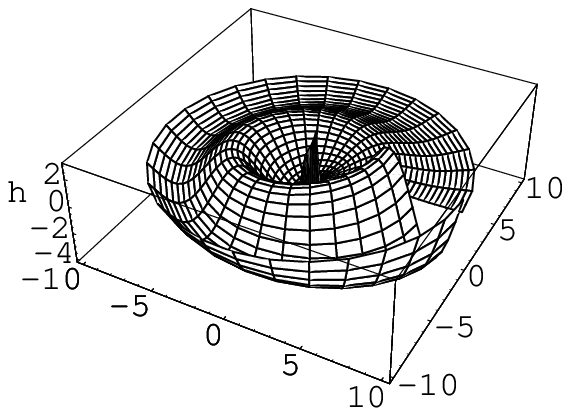}
\end{center}
\caption{fig (1c)}
\end{figure}
\begin{figure}[h]
\begin{center}
\includegraphics[width=60mm,height=60mm]{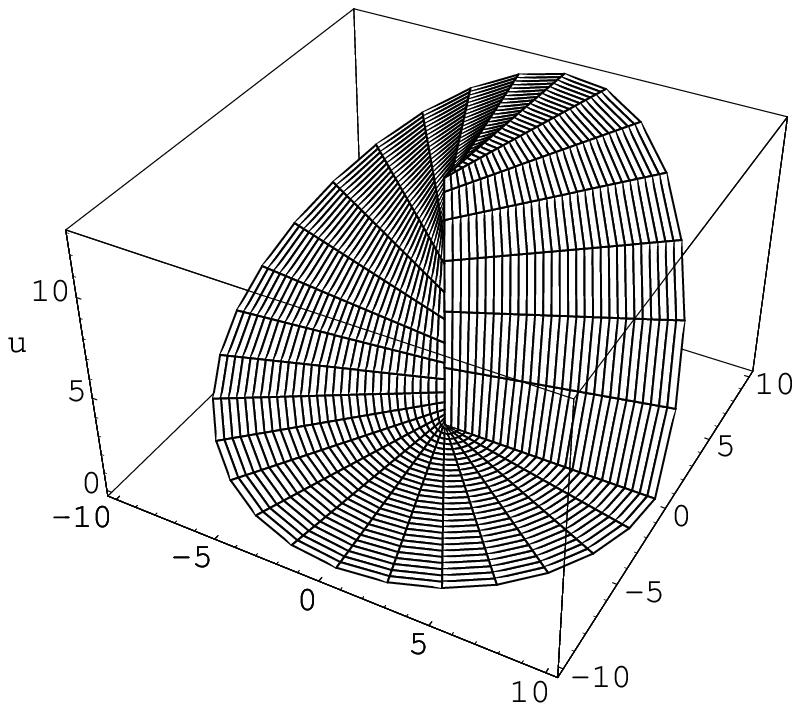}
\end{center}
\caption{fig (2a)}
\end{figure}
\begin{figure}[h]
\begin{center}
\includegraphics[width=60mm,height=60mm]{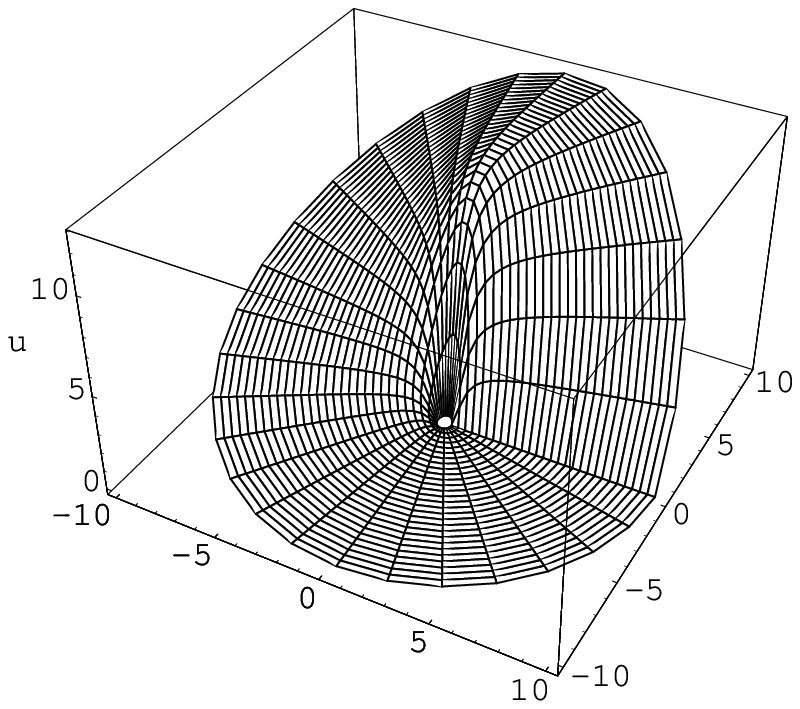}
\end{center}
\caption{fig (2b)}
\end{figure}
\begin{figure}[h]
\begin{center}
\includegraphics[width=60mm,height=60mm]{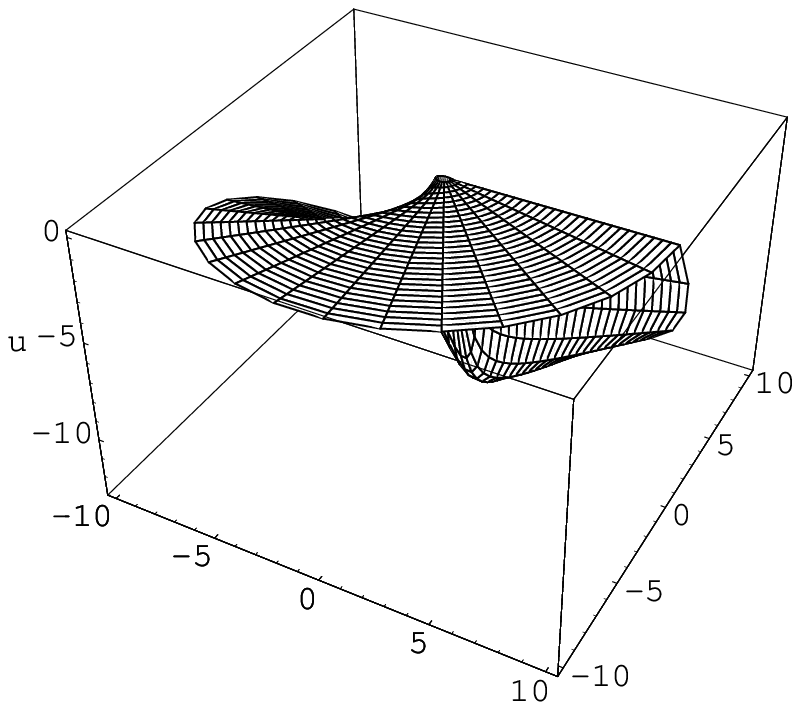}
\end{center}
\caption{fig (2c)}
\end{figure}

\begin{figure}[h]
\begin{center}
\includegraphics[width=60mm,height=60mm]{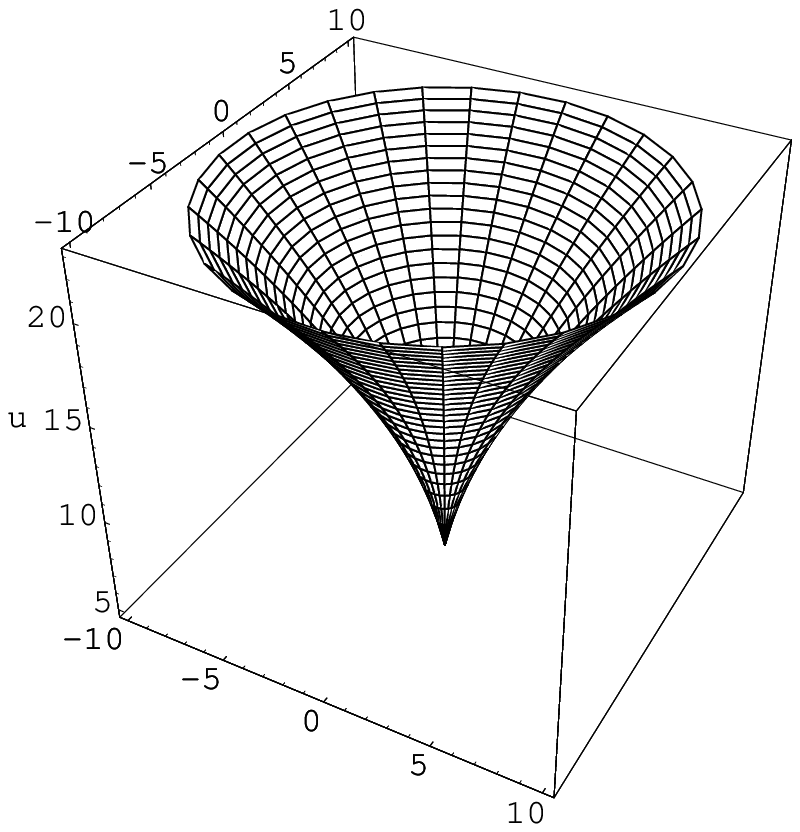}
\end{center}
\caption{fig (3a)}
\end{figure}
\begin{figure}[h]
\begin{center}
\includegraphics[width=60mm,height=60mm]{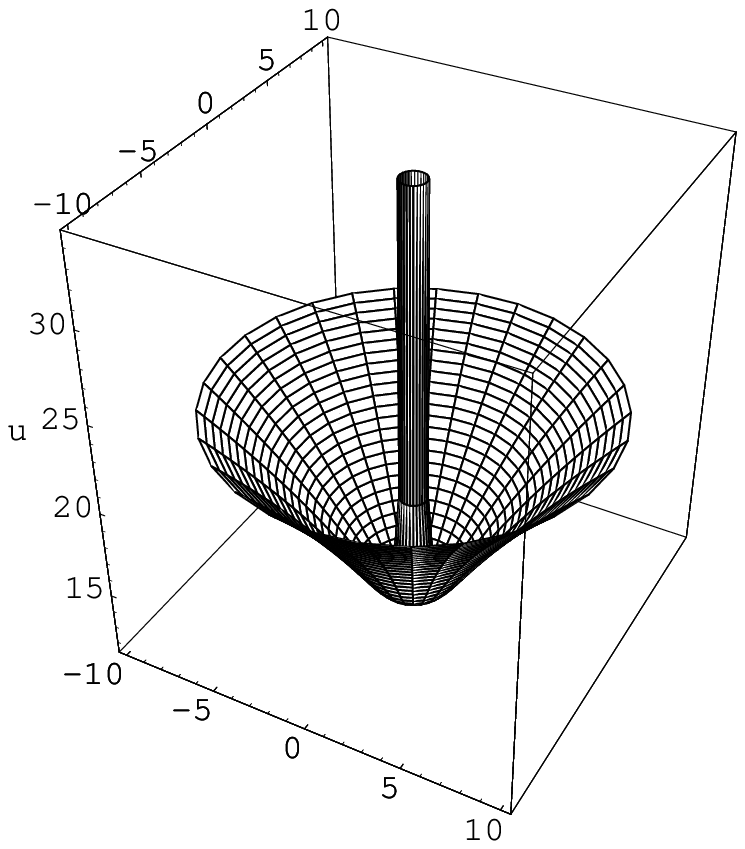}
\end{center}
\caption{fig (3b)}
\end{figure}
\begin{figure}[h]
\begin{center}
\includegraphics[width=60mm,height=60mm]{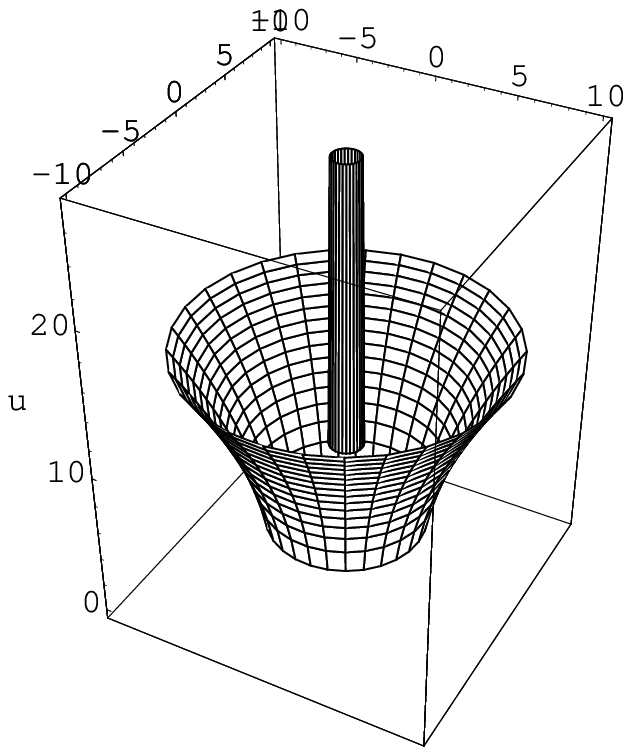}
\end{center}
\caption{fig (3c)}
\end{figure}

\end{document}